\documentclass[twocolumn, 5p, 12pt]{elsarticle}
\usepackage{graphicx}
\usepackage{subfigure}
\usepackage[dvipdfx,cmyk]{xcolor}
\usepackage{amssymb}

\biboptions{sort&compress}

\journal{Physics Letters A}

\usepackage[utf8]{inputenc}
\usepackage[T1]{fontenc}
\usepackage{graphicx}
\usepackage{float}
\usepackage{amssymb}
\usepackage{amsmath}
\usepackage{hyperref}
\usepackage{subfigure}
\usepackage{color}
\usepackage{multirow}
\usepackage{relsize}
\usepackage{url}

\definecolor{mycolora}{rgb}{1.000, 0.000, 0.000}
\definecolor{mycolorb}{rgb}{0.707, 0.707, 0.000}
\definecolor{mycolorc}{rgb}{0.000, 1.000, 0.000}
\definecolor{mycolord}{rgb}{0.000, 0.707, 0.707}
\definecolor{mycolore}{rgb}{0.000, 0.000, 1.000}
\newcommand\myusecolora[1]{{\bf \color{mycolora} #1}}
\newcommand\myusecolorb[1]{{\bf \color{mycolorb} #1}}
\newcommand\myusecolorc[1]{{\bf \color{mycolorc} #1}}
\newcommand\myusecolord[1]{{\bf \color{mycolord} #1}}
\newcommand\myusecolore[1]{{\bf \color{mycolore} #1}}

\newcommand\GR{{G_{\rm R}}}
\newcommand\Grr{{G_{rr}}}

\newcommand\Gqr{{G_{\rm qr}}}
\newcommand\Gpr{{G_{\rm pr}}}
\newcommand\Gqrd{{G_{\rm qrd}}}
\newcommand\Gqrnd{{G_{\rm qrnd}}}
\newcommand\Wqr{{W_{\rm qr}}}
\newcommand\Wpr{{W_{\rm pr}}}
\newcommand\Wrr{{W_{\rm rr}}}
\newcommand\Wqrnd{{W_{\rm qrnd}}}
\newcommand\WR{{W_{\rm R}}}

\begin{document}

\begin{frontmatter}
\title{What is the central bank of Wikipedia?}

\author[denis]{Denis Demidov}
\ead{dennis.demidov@gmail.com}

\address[denis]{Kazan Branch of Joint Supercomputer Center, 
Scientific Research Institute of System Analysis, \\
Russian Academy of Sciences, 42011 Kazan, Russia}

\author[klaus]{Klaus M.~Frahm}
\ead{frahm@irsamc.ups-tlse.fr}

\address[klaus]{Laboratoire de Physique Th\'{e}orique, IRSAMC, 
Universit\'{e} de Toulouse, 
CNRS, UPS, 31062 Toulouse, France}

\author[dima]{Dima L.~Shepelyansky}
\ead{dima@irsamc.ups-tlse.fr}

\address[dima]{Laboratoire de Physique Th\'{e}orique, IRSAMC, 
Universit\'{e} de Toulouse, CNRS, UPS, 31062 Toulouse, France}

\begin{abstract}
We analyze the influence and interactions
of 60 largest world banks for 195 world countries
using the reduced Google matrix algorithm for
the English Wikipedia network
with 5 416 537 articles. While the top
asset rank positions are taken by the banks of
China, with China Industrial and Commercial Bank of China
at the first place, we show that the network
influence is dominated by USA banks with
Goldman Sachs being the central bank.
We determine the network structure of interactions of banks and
countries and PageRank sensitivity of countries to
selected banks. We also present GPU oriented
code which significantly accelerates the numerical
computations of reduced Google matrix.
\end{abstract}




\end{frontmatter}

\section{Introduction}
\label{sec:1}

The world influence of the bank system was clearly demonstrated by the 
financial crisis of 2007-2008 whose impact appeared on financial, 
economic and political 
levels of many world countries (see e.g. \cite{fcrisiswiki,fcrisisguardian}).
The world wide contamination propagation of this crisis 
pushed forward the scientific analysis of bank networks
which allows to detect interconnections of financial flows between banks
\cite{garratt2010,hale2012,minoiu2013,craig2014}.
For a bank network it is important to know what is the central bank
since it may propagate its influence via the   financial network
to many other banks all over the world.
On a first glance one could expect that a bank with 
the biggest total asset will be the most influential one.
The list of the largest 60 world banks, ranked by  their total
assets, is available at \cite{wikibankrank} on the basis of the 
S\&P Global Market Intelligence report 2018 \cite{spreport}.
The top position of rank index $K_a=1$
is taken by  Industrial and Commercial Bank (ICB) of China.
The distribution of these 60 banks over world countries 
is given in Table~\ref{tab1}
and their names and AssetRank index $K_a$ are given in Table~\ref{tab2}.
The first 4 positions are taken by banks of China
and the first US bank appears only at 6th position
taken by  JPMorgan Chase \& Co. The 5th position is taken by
Mitsubishi UFJ Financial Group from Japan.

However, in the network of nodes the influence is determined
by the links between nodes. 
Indeed, the PageRank algorithm was proposed by Brin and Page in 1998 
\cite{brin}
with the aim to rank the nodes of the World Wide Web (WWW).
The method uses the construction of the Google matrix of Markov chain 
transitions between network nodes connected by directed links.  
This efficient algorithm became the foundation of the Google search 
engine \cite{brin,meyer}.  
The efficiency of this approach has been demonstrated for various types
of directed networks including WWW, Wikipedia networks \cite{rmp2015}
and the world trade networks \cite{wtn1,wtn2}. The PageRank probability vector
being the eigenvector of the largest eigenvalue of the Google matrix 
determines the most important and influential nodes
on the network \cite{brin,meyer}.

\begin{table}
\begin{center}
\caption{List of 15 countries associated with the 60 richest banks 
from \cite{wikibankrank} 
with $n_b$ being the number of banks for each country appearing in 
the bank list. The second column represents  $\alpha_2$ ISO 3166-1
country codes. 
The colors mark five groups with group index $n_g$: 
US and Canada ($n_g=1$, red), 
Switzerland and UK ($n_g=2$, olive), EU countries without UK 
($n_g=3$, green), India, Australia and Japan ($n_g=4$, cyan) and 
China ($n_g=5$, blue).}
\label{tab1}
{\relsize{-2}
\vspace{-0.3cm}
\begin{tabular}{llrr}
\hline
Country & $\alpha_2 $ & $n_b$ & $n_g$ \\
\hline
\myusecolora{United States} & \myusecolora{US} & 6 & 1 \\
\myusecolorb{United Kingdom} & \myusecolorb{UK} & 5 & 2 \\
\myusecolorc{Germany} & \myusecolorc{DE} & 3 & 3 \\
\myusecolorb{Switzerland} & \myusecolorb{CH} & 2 & 2 \\
\myusecolorc{France} & \myusecolorc{FR} & 5 & 3 \\
\myusecolora{Canada} & \myusecolora{CA} & 5 & 1 \\
\myusecolorc{Netherlands} & \myusecolorc{NL} & 2 & 3 \\
\myusecolorc{Italy} & \myusecolorc{IT} & 3 & 3 \\
\myusecolord{India} & \myusecolord{IN} & 1 & 4 \\
\myusecolord{Australia} & \myusecolord{AU} & 4 & 4 \\
\myusecolorc{Spain} & \myusecolorc{ES} & 2 & 3 \\
\myusecolore{China} & \myusecolore{CN} & 13 & 5 \\
\myusecolorc{Finland} & \myusecolorc{FI} & 1 & 3 \\
\myusecolord{Japan} & \myusecolord{JP} & 7 & 4 \\
\myusecolorc{Denmark} & \myusecolorc{DK} & 1 & 3 \\
\hline
\end{tabular}

}
\end{center}
\end{table}

In \cite{garratt2010} it was proposed to use the PageRank vector
and to consider that the central bank is the one
at the top of PageRank probability with PageRank index $K_b=1$.
The analysis was done for the Canadian Large Value Transfer System
with 14 banks. While this study presents an interesting 
approach to the analysis of bank influence it is clear that
the network of 14 Canadian banks is too small
to make any conclusion for the world influence of banks.
Also it is highly difficult to obtain information on
monetary transfers between banks on the world scale
due to the high secrecy of bank operations.
At the same time it is clear that
the monetary transfers do not provide
the whole information about the real influence
of banks since it includes also 
political, social, historical and other
types of links and relations which are not
directly expressed in money. 

\begin{table}
\begin{center}
\caption{Table of 60 most important banks. $K_b$ is the relative rank number 
obtained from the Wikipedia PageRank of the 
Wikipedia article of each bank and $K_a$ is the rank number due to 
the total value of assets of the Bank. 
The colors and $\alpha_2$ are as 
in Table~\ref{tab1}. The name ``ICB China'' is an abbreviation of 
``Industrial and Commercial Bank of China''. 
}
\label{tab2}
{\relsize{-2}
\vspace{-0.3cm}
\begin{tabular}{rllr}
\hline
$K_b$ & Bank & $\alpha_2 $  & $K_a$\\
\hline
1 & \myusecolora{Goldman Sachs} & \myusecolora{US} & 35\\
2 & \myusecolora{Citigroup} & \myusecolora{US} & 13\\
3 & \myusecolora{Bank of America} & \myusecolora{US} & 9\\
4 & \myusecolorb{HSBC} & \myusecolorb{UK} & 7\\
5 & \myusecolora{JPMorgan Chase} & \myusecolora{US} & 6\\
6 & \myusecolorb{Barclays} & \myusecolorb{UK} & 18\\
7 & \myusecolorc{Deutsche Bank} & \myusecolorc{DE} & 15\\
8 & \myusecolora{Morgan Stanley} & \myusecolora{US} & 38\\
9 & \myusecolorb{UBS} & \myusecolorb{CH} & 34\\
10 & \myusecolora{Wells Fargo} & \myusecolora{US} & 11\\
11 & \myusecolorb{Credit Suisse} & \myusecolorb{CH} & 40\\
12 & \myusecolorc{BNP Paribas} & \myusecolorc{FR} & 8\\
13 & \myusecolora{Royal Bank of Canada} & \myusecolora{CA} & 24\\
14 & \myusecolorc{Société Générale} & \myusecolorc{FR} & 19\\
15 & \myusecolorb{Standard Chartered} & \myusecolorb{UK} & 48\\
16 & \myusecolorc{ING Group} & \myusecolorc{NL} & 26\\
17 & \myusecolora{Bank of Montreal} & \myusecolora{CA} & 52\\
18 & \myusecolorc{UniCredit} & \myusecolorc{IT} & 28\\
19 & \myusecolora{Scotiabank} & \myusecolora{CA} & 42\\
20 & \myusecolord{State Bank of India} & \myusecolord{IN} & 55\\
21 & \myusecolord{Commonwealth Bank} & \myusecolord{AU} & 43\\
22 & \myusecolorc{Banco Santander} & \myusecolorc{ES} & 16\\
23 & \myusecolorc{Crédit Agricole} & \myusecolorc{FR} & 10\\
24 & \myusecolord{National Australia Bank} & \myusecolord{AU} & 50\\
25 & \myusecolord{Westpac} & \myusecolord{AU} & 47\\
26 & \myusecolord{Australia and New Zealand BG} & \myusecolord{AU} & 45\\
27 & \myusecolorb{Lloyds Banking Group} & \myusecolorb{UK} & 23\\
28 & \myusecolore{ICB China} & \myusecolore{CN} & 1\\
29 & \myusecolorc{Nordea} & \myusecolorc{FI} & 46\\
30 & \myusecolord{Mitsubishi UFJ Financial Group} & \myusecolord{JP} & 5\\
31 & \myusecolorc{Banco Bilbao Vizcaya Argentaria} & \myusecolorc{ES} & 41\\
32 & \myusecolore{Bank of China} & \myusecolore{CN} & 4\\
33 & \myusecolora{Toronto-Dominion Bank} & \myusecolora{CA} & 25\\
34 & \myusecolora{Canadian Imperial B. of Commerce} & \myusecolora{CA} & 59\\
35 & \myusecolorb{Royal Bank of Scotland Group} & \myusecolorb{UK} & 29\\
36 & \myusecolord{Mizuho Financial Group} & \myusecolord{JP} & 17\\
37 & \myusecolorc{Rabobank} & \myusecolorc{NL} & 44\\
38 & \myusecolorc{Commerzbank} & \myusecolorc{DE} & 54\\
39 & \myusecolorc{Intesa Sanpaolo} & \myusecolorc{IT} & 32\\
40 & \myusecolore{China Construction Bank} & \myusecolore{CN} & 2\\
41 & \myusecolore{Agricultural Bank of China} & \myusecolore{CN} & 3\\
42 & \myusecolorc{Danske Bank} & \myusecolorc{DK} & 53\\
43 & \myusecolord{Sumitomo Mitsui Financial Group} & \myusecolord{JP} & 14\\
44 & \myusecolorc{Groupe BPCE} & \myusecolorc{FR} & 20\\
45 & \myusecolorc{Cassa Depositi e Prestiti} & \myusecolorc{IT} & 57\\
46 & \myusecolorc{DZ Bank} & \myusecolorc{DE} & 51\\
47 & \myusecolore{Bank of Communications} & \myusecolore{CN} & 21\\
48 & \myusecolord{Resona Holdings} & \myusecolord{JP} & 56\\
49 & \myusecolord{Sumitomo Mitsui Trust Holdings} & \myusecolord{JP} & 60\\
50 & \myusecolorc{Crédit Mutuel} & \myusecolorc{FR} & 39\\
51 & \myusecolore{China Merchants Bank} & \myusecolore{CN} & 31\\
52 & \myusecolore{China Minsheng Bank} & \myusecolore{CN} & 36\\
53 & \myusecolord{Japan Post Bank} & \myusecolord{JP} & 12\\
54 & \myusecolord{Norinchukin Bank} & \myusecolord{JP} & 27\\
55 & \myusecolore{Ping An Bank} & \myusecolore{CN} & 58\\
56 & \myusecolore{China CITIC Bank} & \myusecolore{CN} & 37\\
57 & \myusecolore{Shanghai Pudong Development B.} & \myusecolore{CN} & 33\\
58 & \myusecolore{Industrial Bank (China)} & \myusecolore{CN} & 30\\
59 & \myusecolore{Postal Savings Bank of China} & \myusecolore{CN} & 22\\
60 & \myusecolore{China Everbright Bank} & \myusecolore{CN} & 49\\
\hline
\end{tabular}

}
\end{center}
\end{table}

Due to these reasons we use another approach based on 
the Wikipedia network analysis. Indeed, Wikipedia 
accumulates a great amount of human knowledge and
supersedes other encyclopedia such as 
Encyclopedia Britannica \cite{gileswiki}.
The academic analysis of information contained in Wikipedia 
finds more and more applications as reviewed in \cite{reagle,nielsen}.
Scientific analysis shows that the quality of Wikipedia articles
is growing \cite{wikiquality}. 

Thus we take here the whole network
of English Wikipedia dated by May 2017 \cite{24wiki2017} 
containing $N=5 416 537$ articles (nodes)
with $N_\ell = 122 232 932$ directed links between them generated by
citations of articles in other articles. 
The global Google matrix of this directed network
is created by the standard rules described in \cite{brin,meyer,rmp2015}.
To analyze the interactions and influence of banks
we select $N_b=60$ largest banks from \cite{wikibankrank},
given also in Table~\ref{tab2}, and $N_c=195$ world countries.
Then we perform the reduced Google matrix (REGOMAX) analysis
\cite{greduced,politwiki}
of these selected $N_r=N_b+N_c=255$ nodes.
The REGOMAX approach allows to determine 
interactions between banks and countries
by direct and indirect links between these selected nodes. 
The indirect links take into account all possible pathways from one node to 
another one via the global network of 5 millions of Wikipedia articles.
The efficiency of the REGOMAX method has been demonstrated for various 
examples such as interactions between politicians \cite{politwiki},
countries \cite{wikicountries}, world universities\cite{wrwu2017} and
cancer networks \cite{proteins}.

The results of our analysis show that the central bank of Wikipedia
is not at all ICB China with the largest asset $K_a=1$ but 
Goldman Sachs which has the asset rank $K_a=35$.
We also determine the influence of 60 banks on 195 world countries.

The complete data and additional figures obtained in this work
are available at \cite{wikibanknet} including the list of 195 countries.

We briefly describe the Google matrix methods in Section~\ref{sec:2},
the rank plane of banks is analyzed in Section~\ref{sec:3},
the reduced Google matrix of banks and countries is described in 
Section~\ref{sec:4}, the network structure of friends and followers 
for banks and countries is shown and discussed in Section~\ref{sec:5}, 
the world country sensitivity
to specific banks are obtained in Section~\ref{sec:6}
and the discussion of the results is given in  Section~\ref{sec:7}.

\section{Google matrix methods}
\label{sec:2}

The Google matrix $G$ is constructed from 
the adjacency matrix $A_{ij}$ with elements $1$ 
if article (node) $j$ 
points to  article (node) $i$ and zero otherwise. 
The matrix elements have the standard form 
$G_{ij} = \alpha S_{ij} + (1-\alpha) / N$ \cite{brin,meyer,rmp2015},
where $S$ is the matrix of Markov transitions with elements  
$S_{ij}=A_{ij}/k_{out}(j)$ and $k_{out}(j)=\sum_{i=1}^{N}A_{ij}\neq 0$ 
being the  out-degree of node $j$ 
(number of outgoing links);  $S_{ij}=1/N$ if $j$ 
has no outgoing links (dangling node). 
The parameter $0< \alpha <1$ is the damping factor.
We use the standard value $\alpha=0.85$ \cite{meyer}
noting that for the range $0.5 \leq \alpha \leq 0.95$
the results are not sensitive to $\alpha$ \cite{meyer,rmp2015}. 
For a random surfer, moving from one node to another,
 the probability to jump to any node is  $(1-\alpha)$. 

 The right PageRank eigenvector
of $G$ is the solution of the equation $G P = \lambda P$
for the unit eigenvalue $\lambda=1$. The PageRank 
$P(j)$ values give positive probabilities to find a random surfer 
on a node $j$ ($\sum_j P(j)=1$). 
We order all nodes by decreasing probability $P$ 
numbered by  PageRank index $K=1,2,...N$ with a maximal probability at $K=1$
and minimal at $K=N$. The numerical computation 
of $P(j)$ is done efficiently with the PageRank
algorithm described in \cite{brin,meyer}.

We also consider
the original network with inverted direction of links.
After inversion the Google matrix $G^*$ is constructed via
the same procedure with $G^* P^*= P^*$. The matrix $G^*$  
has its own PageRank vector
$P^*(j)$ called CheiRank \cite{cheirank} (see also \cite{rmp2015}).
Its probability values can be again ordered
in a decreasing order with CheiRank index $K^*$
with highest  $P^*$ at $K^*=1$ and smallest at $K^*=N$.
On average, the high values of $P$ ($P^*$) correspond to nodes
with many ingoing (outgoing) links \cite{meyer,rmp2015}.

The REGOMAX method is described in detail 
in \cite{greduced,politwiki,wikicountries,proteins}. It allows to 
compute efficiently a ``reduced Google matrix'' of size $N_r \times N_r$ 
that captures
the full contributions of direct and indirect pathways happening 
in the full Google matrix between $N_r$ nodes of interest. 
For the selected $N_r$ nodes 
their PageRank probabilities are the same 
as for the global network with $N$ nodes, 
up to a constant multiplicative factor taking into account that 
the sum of PageRank probabilities over $N_r$
nodes is unity.   
The computation of $\GR$ provides 
a decomposition of $\GR$ into matrix components that clearly distinguish 
direct from indirect interactions: 
$\GR = \Grr + \Gpr + \Gqr$ \cite{politwiki}.
Here $\Grr$ is given by the direct links between selected 
$N_r$ nodes in the global $G$ matrix with $N$ nodes. In fact, 
 $\Gpr$ is rather close to 
the matrix in which each column is given by 
the PageRank vector $P_r$, ensuring that PageRank probabilities of $\GR$ are 
the same as for $G$ (up to a constant multiplier).
Hence $\Gpr$ doesn't provide much information about direct 
and indirect links between selected nodes.

\begin{figure}[h!]
\vskip -0.3cm
\begin{center}
\includegraphics[width=0.50\textwidth]{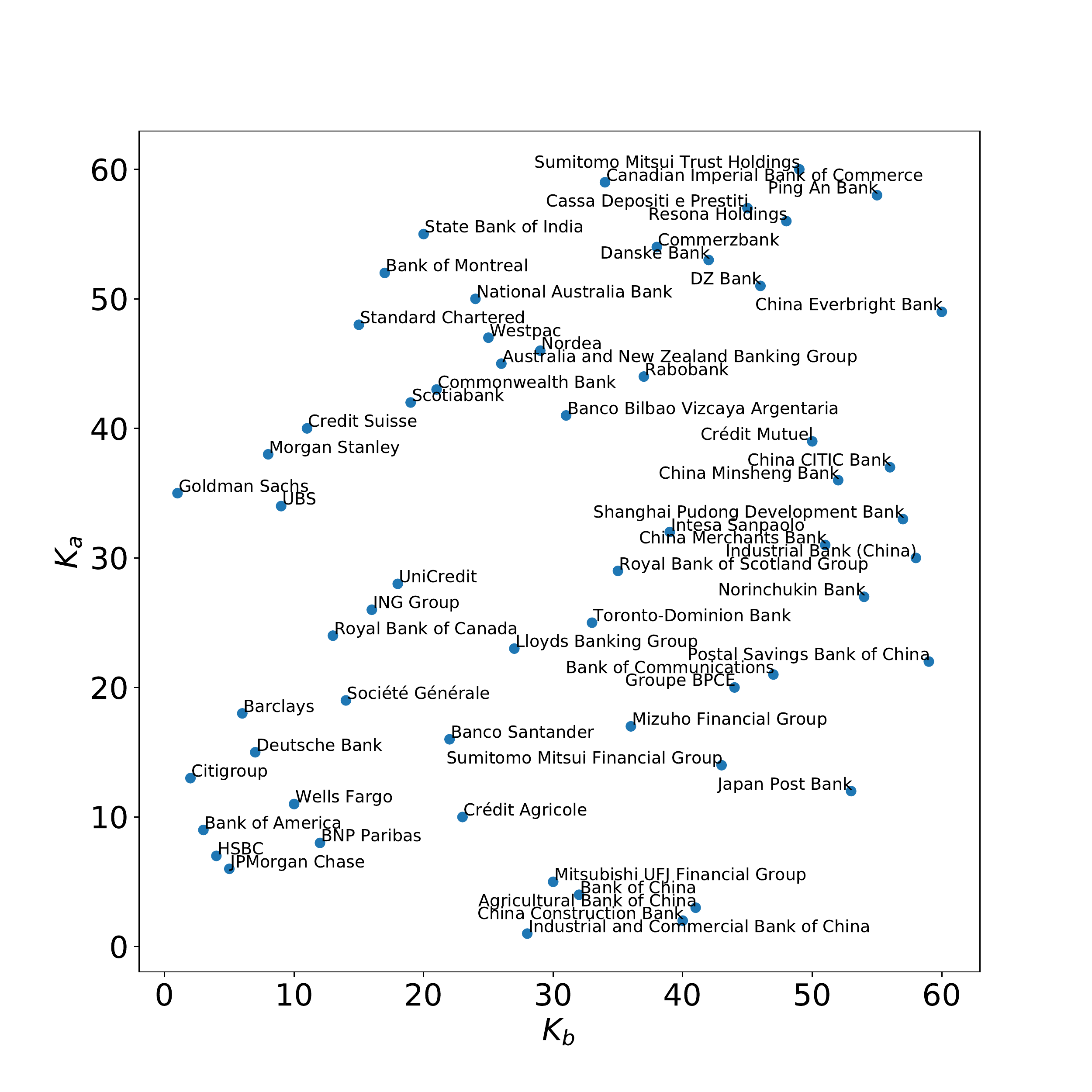}\\
\includegraphics[width=0.50\textwidth]{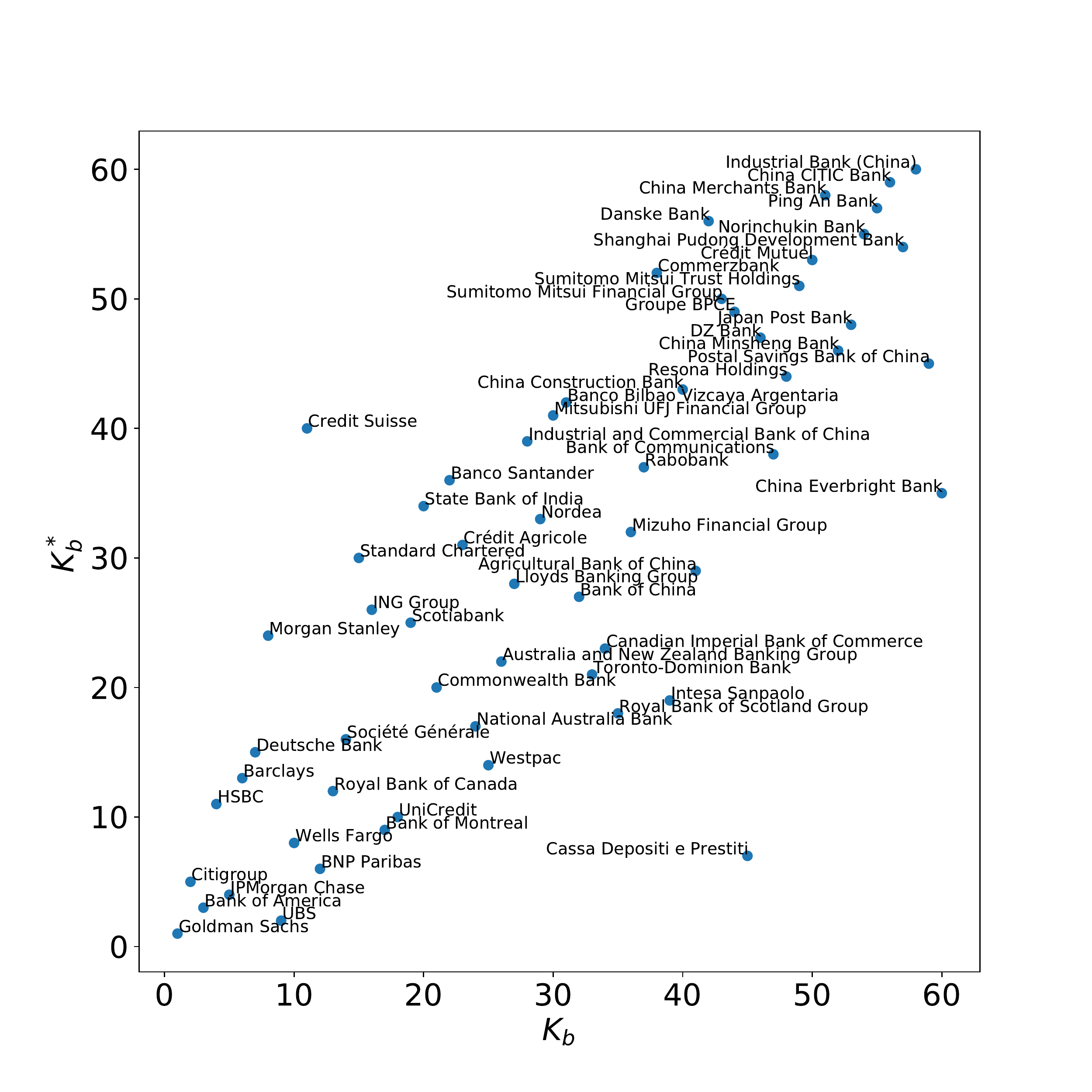}
\caption{Top panel: PageRank-AssetRank plane of 60 banks 
with the world largest assets from \cite{wikibankrank};
bottom panel: PageRank-CheiRank plane of the same banks.
}
\label{fig1}
\end{center}
\end{figure}

The most interesting role is played by $\Gqr$, which takes 
into account all indirect links between
selected nodes appearing due to multiple pathways via 
the global network nodes $N$ (see~\cite{greduced,politwiki}).
The matrix  $\Gqr = \Gqrd + \Gqrnd$ has diagonal ($\Gqrd$)
and non-diagonal ($\Gqrnd$) parts with $\Gqrnd$
 describing indirect interactions between nodes.
The exact formulas for  all three components of $\GR$ are given 
in \cite{greduced,politwiki}.
The numerical computation 
methods of all three components of $\GR$ and numerical REGOMAX code
developed in  \cite{greduced}
are given in  \cite{gitlabklaus}.

In this work we also ported the original REGOMAX code to NVIDIA CUDA technology
using VexCL library~\cite{demidov2013, vexcl}, with two notable modifications.
First, we reordered the global Google matrix using Cuthill-McKee
method~\cite{cuthillmckee} in order to reduce the bandwidth of the matrix, and
thus improve cache locality of the code. This resulted in about 30\% reduction
of the computational time. Second, the computation of the reduced matrix was
performed in batches of 8-20 columns, which allowed us to further improve the
performance of the method by another 50\%.  The computations were run at the 
OLYMPE CALMIP cluster \cite{olympe} using NVIDIA Tesla V100 GPUs. 
Overall, we managed
to achieve an approximately 8x speedup with respect to the original OpenMP
implementation which was performed on a dual socket Intel(R) Xeon(R) 
CPU E5-2620
v2 @ 2.10GHz system ($2\times 6$ cores). The REGOMAX-GPU code is available at
\cite{gitlabdenis}. 

With the matrix  $\GR$ and its components we can analyze 
the PageRank sensitivity
with respect to specific links between $N_r = 255$ nodes. To measure 
the sensitivity of a country $c$ to a bank $b$ we change the matrix element
$G_\mathrm{R}(b \rightarrow c)$ by a factor $(1+\delta)$ with $\delta\ll1$
and renormalize to unity 
the sum of the column elements associated with bank $b$. Then
we compute the logarithmic derivative
of PageRank probability $P(c)$ associated to country $c$:
$D(b \rightarrow c, c) = d \ln P(c)/d\delta$ 
(diagonal sensitivity). 
This approach already demonstrated its efficiency as reported in
\cite{wrwu2017,ieeedis}.

\vspace{-0.4cm}
\section{Rank plane of banks}
\label{sec:3}

We determine the PageRank index of countries and banks for $N_r=255$.
As discussed in \cite{rmp2015,wikicountries}
the top PageRank positions of the global matrix $G$ are taking by countries.
The PageRank local index $K_b$ marking only banks is given in Table~\ref{tab2}.
The distribution of banks on the plane of PageRank $K_b$ and AssetRank $K_a$
is shown in Fig.~\ref{fig1} (top panel). While the top AssetRank
$K_a=1$ is taken by ICB China its PageRank index is only $K_b=28$ 
while the top PageRank position $K_b=1$ belongs to 
Goldman Sachs even if its AssetRank is $K_a=35$.
{\it Thus the central bank of Wikipedia is Goldman Sachs bank}.
While the top 4 positions of AssetRank are taken by banks of China
the top 4 PageRank positions are taken by USA banks (plus HSBC of UK).
Since Wikipedia includes a huge amount of knowledge
accumulated by humanity
we can say the the most strong political and social influence
belongs to USA banks.

In contrast to the PageRank vector, which is reflecting the influence,
the CheiRank vector reflects the communicative features of nodes.
Thus the banks on the top positions of CheiRank index of banks
${K^*}_b$ have the most communicative articles (nodes) among banks.
The distribution of banks of the PageRank-CheiRank plane $(K_b,{K^*}_b)$
is shown in the bottom panel of Fig.~\ref{fig1}.
The top CheiRank positions ${K^*}_b=1,2,3$ are taken by 
Goldman Sachs, UBS, Bank of America,
while  ICB China has ${K^*}_b = 39$
showing very low communicativity of its Wikipedia article.

\begin{figure}[h!]
\begin{center}
\includegraphics[width=0.49\textwidth]{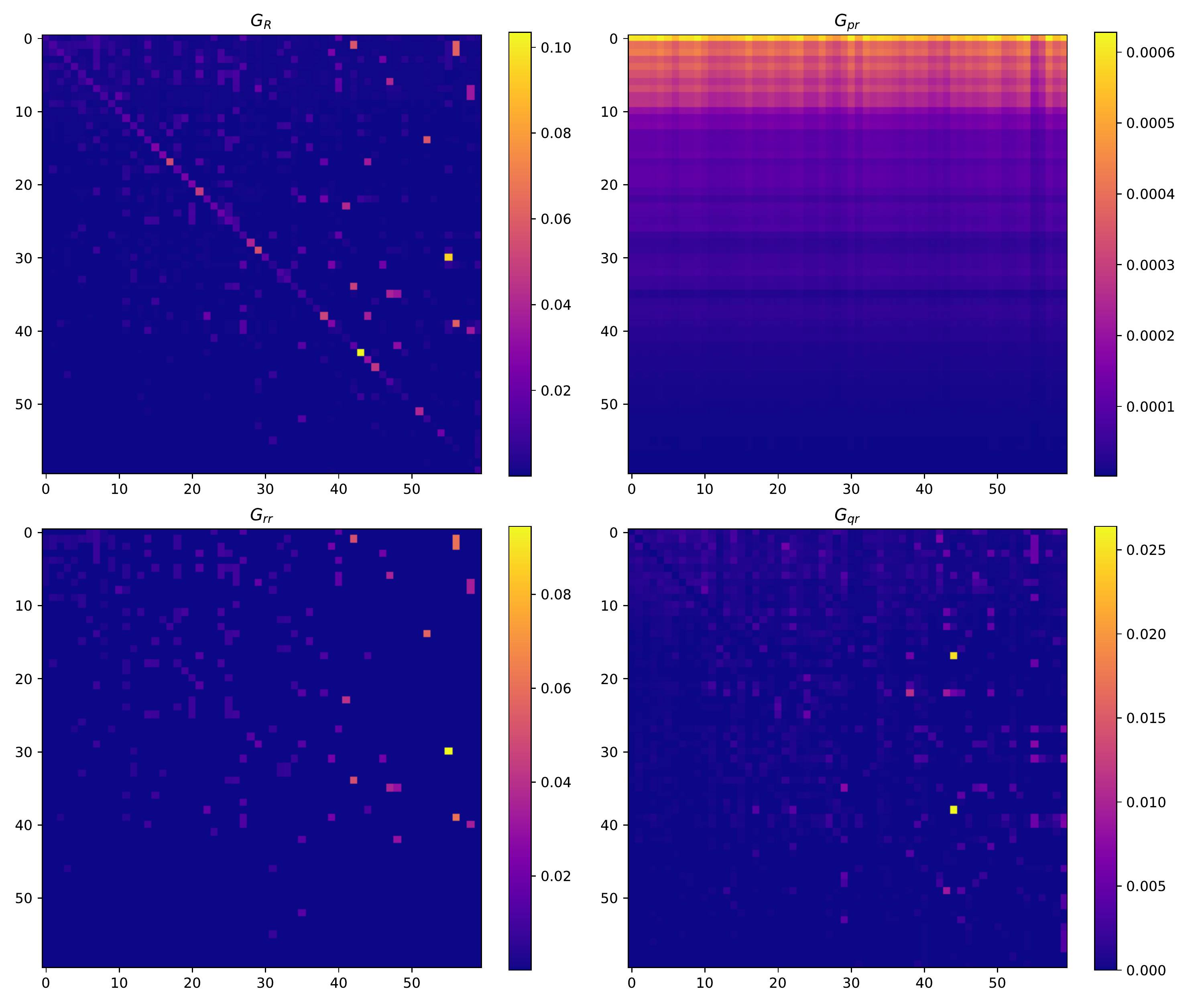}
\caption{Matrix elements of reduced Google matrix 
$\GR$ and its 3 components $\Gpr$, $\Grr$, $\Gqr$
shown for the bank sector $60 \times 60$
of the whole matrix $255 \times 255$.
Bank index on axis is given by the PageRank index $K_b$
of Table~\ref{tab2}.
}
\label{fig2}
\end{center}
\end{figure}

\vspace{-0.4cm}
\section{Reduced Google matrix}
\label{sec:4}

Using the REGOMAX algorithm we determine
the matrix $\GR$ and its 3 components for 
$N_r=255$ banks and countries.
These matrices are available at \cite{wikibanknet}.
We determine the weight of each component as the sum of all 
matrix elements divided by $N_r$ (the weight
of $\GR$ is by definition $\WR=1$).
The weights of the components $\Gpr$, $\Grr$, $\Gqr$ and
$\Gqrnd$ are respectively $\Wpr=0.8912$,
$\Wrr=0.0402$, $\Wqr=0.06859$ and $\Wqrnd=0.04417$.
The weights of components $\Grr$, $\Gqr$ 
are relatively small but as it was discussed above and in 
\cite{politwiki,wrwu2017,wikicountries}
they provide the most interesting information 
on interactions of nodes.

Here, in Fig.~\ref{fig2}
we show the parts of $\GR$ belonging to the 
$60 \times 60$ corner corresponding its bank-bank sector.
The whole matrix $255 \time 255$ is available at \cite{wikibanknet}.
The weight of matrix elements of this $60 \times 60$ sector
is relatively small
with $\WR=0.0800$ and $\Wpr=0.00598$,
$\Wrr=0.03567$, $\Wqr=0.03843$ and $\Wqrnd=0.02267$.
Thus we see that this sector gives only a relatively small 
contribution for the full matrix of banks and countries 
of size $255 \times 255$. However,  this $60 \times 60$ sector
describes important interactions between banks.
Here we see that there is for example a strong indirect link 
in $\Gqr$ from the Postal Savings Bank of China ($K_b=59$) to Goldman Sachs 
($K_b=1$) with matrix element 
$3.74\times 10^{-3}$ while in $\Grr$ the same link is nearly absent 
with matrix element $2.77\times 10^{-8}=(1-\alpha)/N$, only non-zero due to 
the minimal damping factor contribution.

\vspace{-0.4cm}
\section{Networks of banks and countries}
\label{sec:5}

\begin{figure}[h!]
\begin{center}
\includegraphics[width=0.46\textwidth]{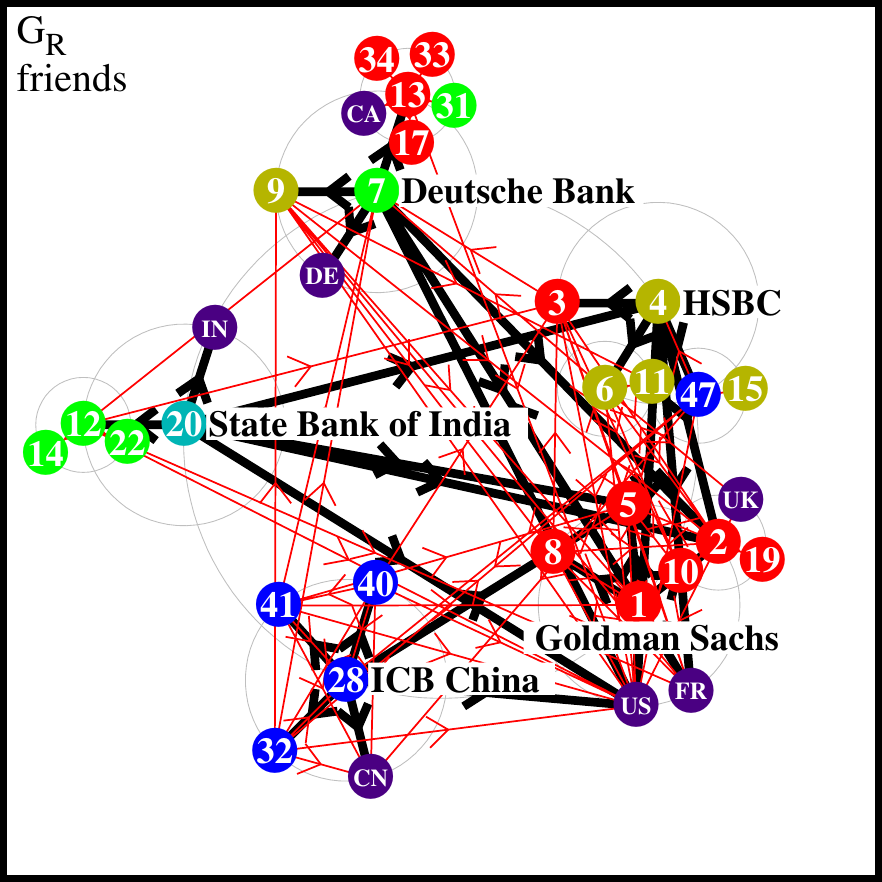}\\
\includegraphics[width=0.46\textwidth]{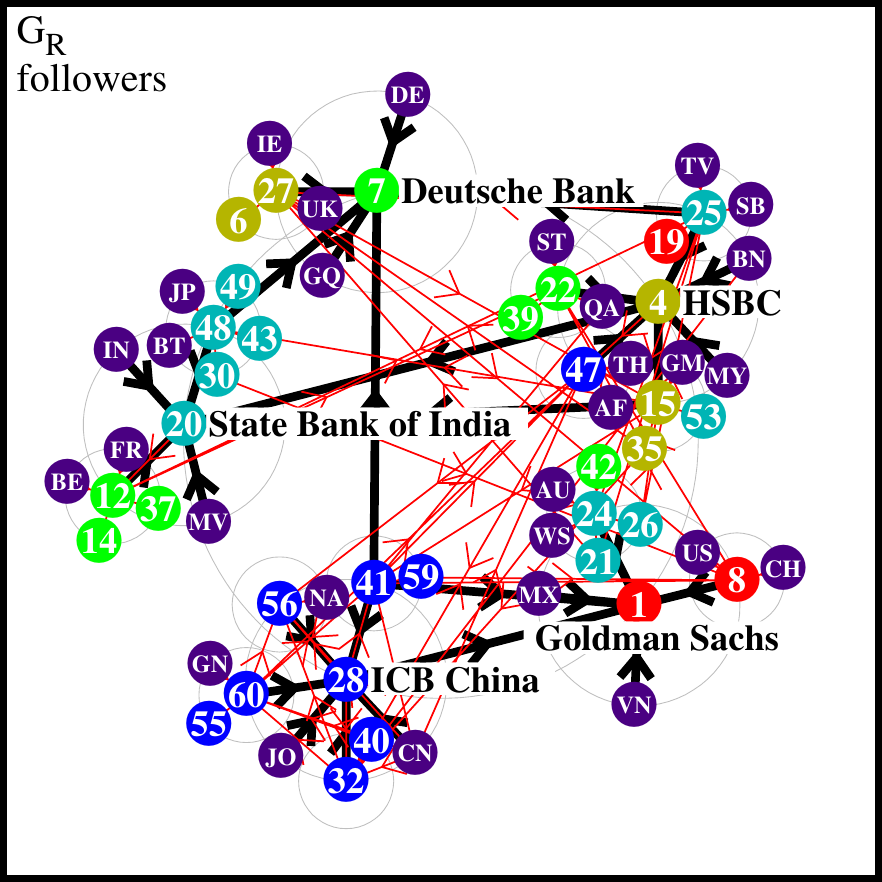}
\caption{Network of friends (top) and followers (bottom)
from $\GR$ matrix of 60 banks and 195 countries;
2 depth levels are shown; links starting from (for friends) or arriving to 
(for followers) level 0 nodes are shown by thick black arrows and similarly 
links starting from/arriving to level 1 nodes are shown by thin red arrows; 
bank nodes are 
marked by their PageRank index $K_b$, given in Table~\ref{tab2}, and 
their group color defined in Table~\ref{tab1};
countries are marked by their $\alpha_2$ ISO 3166-1 country code 
and the color indigo. 
}
\label{fig3}
\end{center}
\end{figure}

At present the network presentation
of link interactions between nodes gained a significant 
popularity due to its compact presentation of interactions
(see e.g. \cite{dorogovtsev}). In directed networks
we will say that a node $j$ has a friend $i$
if $j$ points to $i$ and $i$ has a follower $j$.
Such a description has been used for interactions between politicians
\cite{politwiki}, countries \cite{wikicountries}
and interactions between infectious diseases and countries  \cite{ieeedis}.

We construct similar types of networks from our reduced Google matrix 
$\GR$ of size  $255 \times 255$  for all 60 banks and 195 countries. 
For a better visibility we attribute all 60 banks to $5$ groups according 
to geographical and historical links between countries and banks as 
it it shown in Tables~\ref{tab1},~\ref{tab2}.
The top PageRank banks of each group represent the central group nodes
being Goldman Sachs (red for North America), HSBC (olive for Europe non-EU), 
Deutsche Bank (green for Europe EU)), State Bank of India (cyan for India, 
Japan and Australia ), and ICB China (blue for China). These 
5 top nodes define the set of level 0 nodes. Assuming that the network 
has already been constructed up to level $j$ nodes (with integer $j\ge 0$) we 
determine for each level $j$ bank node 4 bank 
friends (followers) with 4 largest bank matrix elements of $G_R$ in the same 
column (row). If such a friend (follower) is not yet present in the network of 
levels up to $j$ we add it to the network as level $(j+1)$ node. 
Furthermore, we also add in the same way (up to) 2 country friends 
(followers) with 2 largest country matrix elements of $G_R$ in the same 
column (row) for each level $j$ bank node (we do not select followers or 
friends for country nodes once they appear in the network). 

The resulting friend (follower) network is shown in top (bottom) panel of 
Fig.~\ref{fig3} for $G_R$ and maximal depth at level 2. 
The links to friends (from followers) are drawn as arrows 
(thick black line if $j=0$ or thin red line if $j=1$ in the above scheme). 
The nodes are drawn on gray circles of three different sizes for the 
levels 0, 1 and 2. A new (bank) node is attributed to the circle of the 
bank of the same group/color if there is a link. Otherwise it is 
attributed to the circle of the first 
(in $K_b$ order) bank of the previous level to which a link exists. 

We see that in the friend network of $G_R$ there is a large cluster
of USA banks located around Goldman Sachs with the two friend countries
US, FR. The friends of ICB China are three other Chinese banks, one US bank 
and the two countries CN and US. The bank friends of the State Bank of India 
are two European and two US banks together with IN and US as country friends. 
The bank HSBC has two US bank friends, one other UK bank friend and a CN bank 
friend together with US and FR as country friends. The bank friends of 
the Deutsche Bank are from Switzerland (1 bank) and US (3 banks) and its 
country friends are DE and US. 

In the follower network of $G_R$, 
we observe a lot of countries, quite low in the PageRank order for countries, 
which are level 1 or level 2 followers and do not have a bank of their own in 
the list of top 60 banks (i.e. countries which do not appear in 
Tables~\ref{tab1} or \ref{tab2}). The followers of ICB China are other CN 
Banks (exclusively on level 1 and mostly on level 2) and its country followers 
are CN and JO. For the other four top banks the main bank followers are 
quite often from banks of other groups/colors. For the Deutsche Bank and 
the State Bank of India, the corresponding country (DE or IN) is also a 
country follower but the other country follower is a small country 
(GQ or MV) which for GQ is even from a very different geographical 
region. For Goldman Sachs and HSBC both country followers (MX, VN or MY, BN) 
are different from the country to which they belong. Certain other ``small'' 
countries are level 2 followers of level 1 banks. It seems that many countries 
with no powerful bank of their own depend on or are at least related to one or 
several big banks outside or in their region (for example MV for State Bank of 
India or MX for Goldman Sachs). 

Networks built for other cases: other $\GR$ components including $\Grr+\Gqr$, 
higher levels $j>2$ and also the case of a pure bank network 
(without countries) are given in \cite{wikibanknet}. 


\vspace{-0.4cm}
\section{World bank influence on countries}
\label{sec:6}

In addition to the network structure on interactions of banks and
countries, shown in Fig.~\ref{fig3},
we can obtain a more direct and detailed
characterization of bank influence on world countries
via the sensitivity $D(b \rightarrow c, c)$ described
in Section~\ref{sec:2}.

\begin{figure}[h]
\begin{center}
\includegraphics[width=0.50\textwidth]{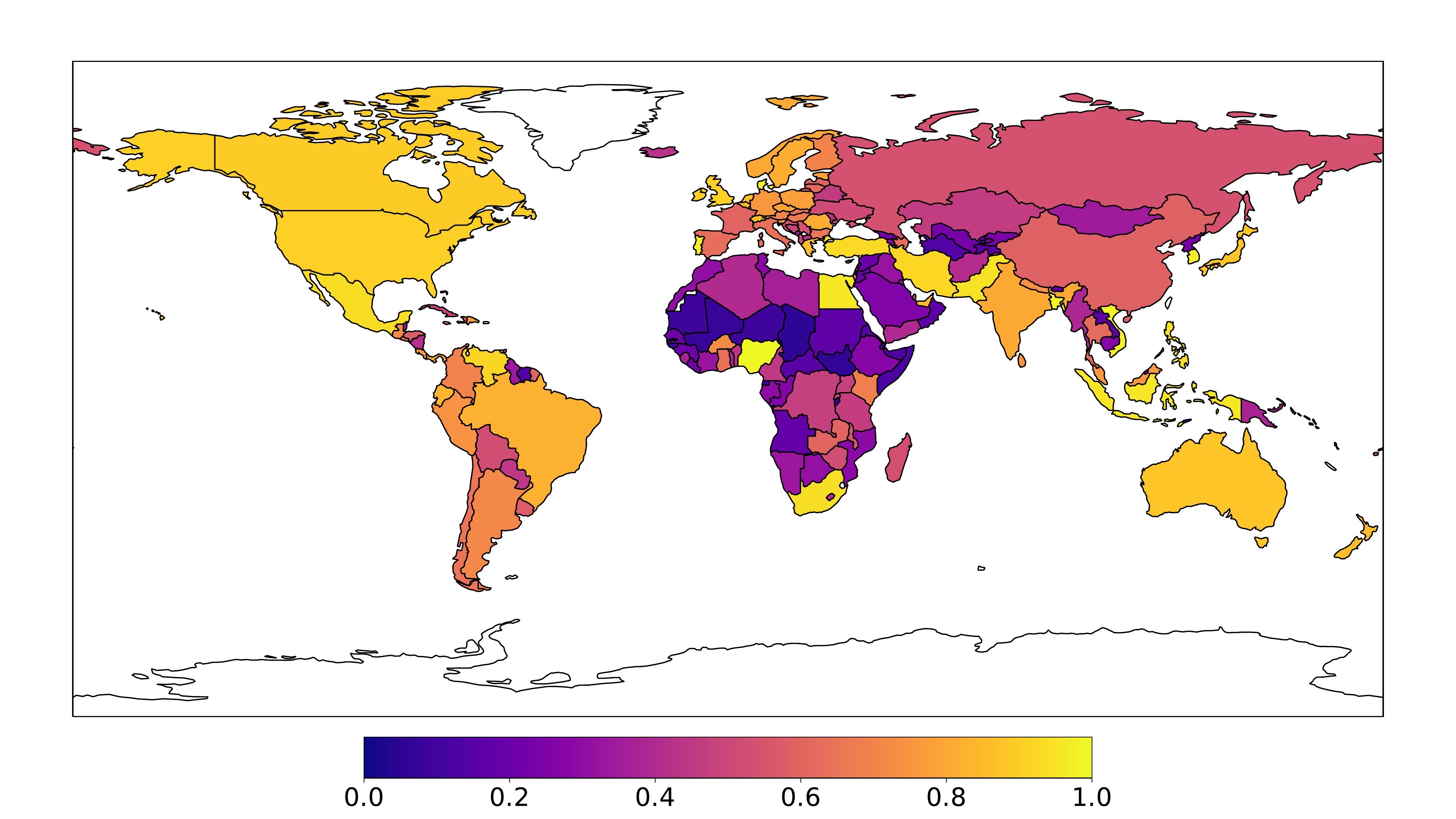}
\caption{Sensitivity $D$ of world countries to Goldman Sachs 
shown by color changing from zero level (blue color for
$D=1.03 \times 10^{-3}$) to maximal unity level 
(yellow color for $D = 5.30 \times 10^{-4}$).
}
\label{fig4}
\end{center}
\end{figure}

Thus in Fig.~\ref{fig4} we show the influence of Goldman Sachs 
on the world countries expressed by the sensitivity $D$
presented on the world map. The strongest sensitivity  is for
Nigeria ($D = 1.03 \times 10^{-3}$) followed by Bangladesh, Vietnam,
Denmark and S.Korea. For UK and USA we have 
$D = 6.7 \times 10^{-4}$ and the minimal sensitivity
$D= 5.3 \times 10^{-4}$ is for Chad. On the world map we see also 
a significant sensitivity for Portugal, Egypt, Indonesia, Pakistan,
Philippines, South Africa, Mexico taking positions from 6 to 12
from the most sensitive country of Nigeria.
These countries belongs to the ``Next Eleven'' list 
marked by Goldman Sachs for countries with
high macroeconomic stability, political maturity, openness of 
trade and investment policies and quality of education as criteria
\cite{goldmanwiki}.

\begin{figure}[h]
\begin{center}
\includegraphics[width=0.50\textwidth]{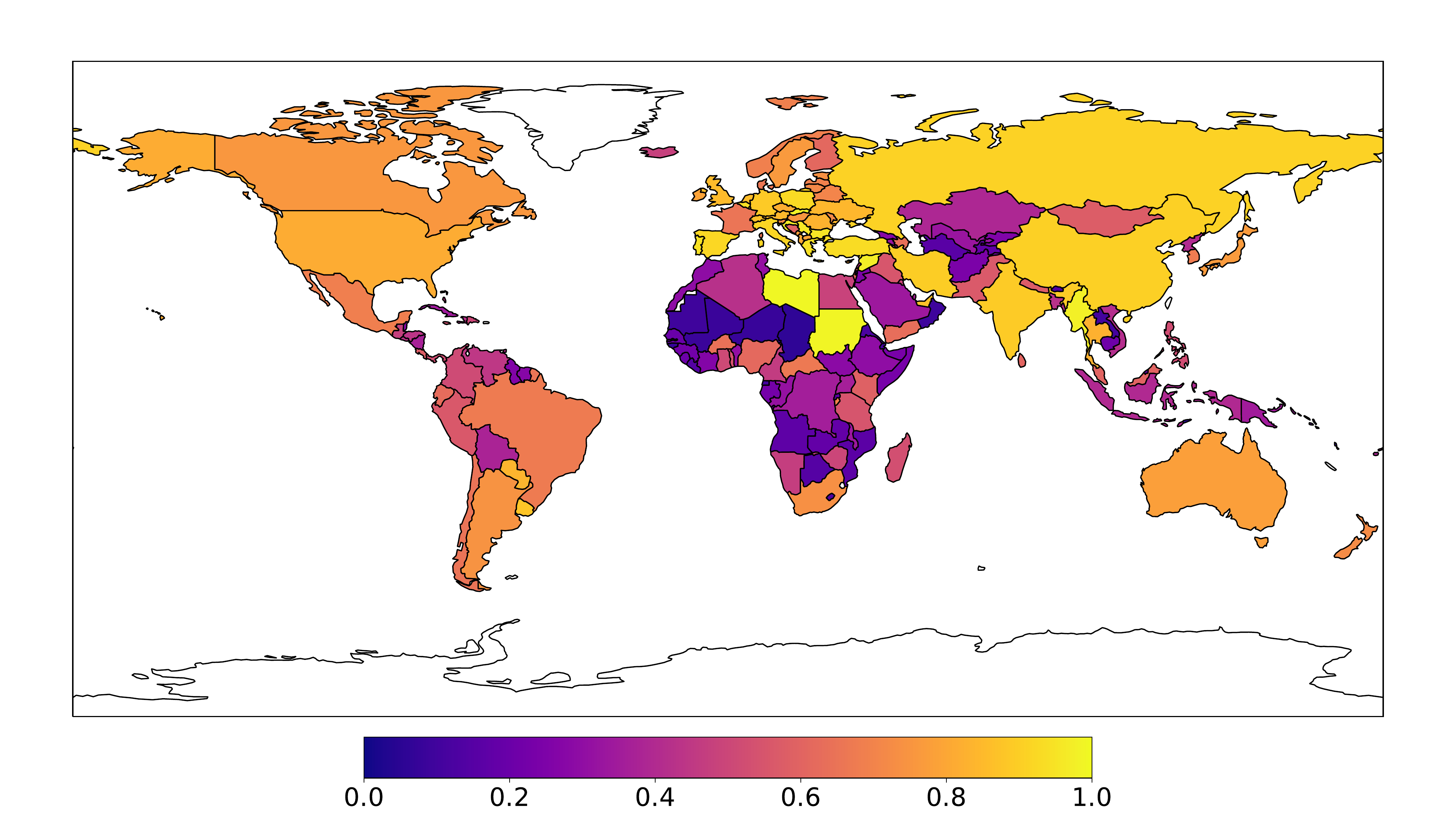}
\caption{Same as in Fig.~\ref{fig4}
for sensitivity to Deutsche Bank
with $D=1.25 \times 10^{-3}$ (blue for zero)
and $D=2.48 \times 10^{-4}$ (yellow for unity).
}
\label{fig5}
\end{center}
\end{figure}

The world influence of  Deutsche Bank is shown
in Fig.~\ref{fig5}.  The highest sensitivity is for Libya
($D=1.25 \times 10^{-3}$) even if there is no direct link
from the wiki-article of Deutsche Bank to Libya. Thus this sensitivity
appears due to indirect links between this country and Deutsche Bank.
Libya is followed by Sudan and Slovakia.
The world map shows that the influence of Deutsche Bank
propagates mainly to East Europe, Turkey, Russia and China.
We note that there is no direct link from  Deutsche Bank
to Russia, but there are links with VTB Bank in Russia
that generates indirect links between Russia and Deutsche Bank.
The lowest sensitivity is for 
East Timor $(D=2.48 \times 10^{-4}$).

\begin{figure}[h]
\begin{center}
\includegraphics[width=0.50\textwidth]{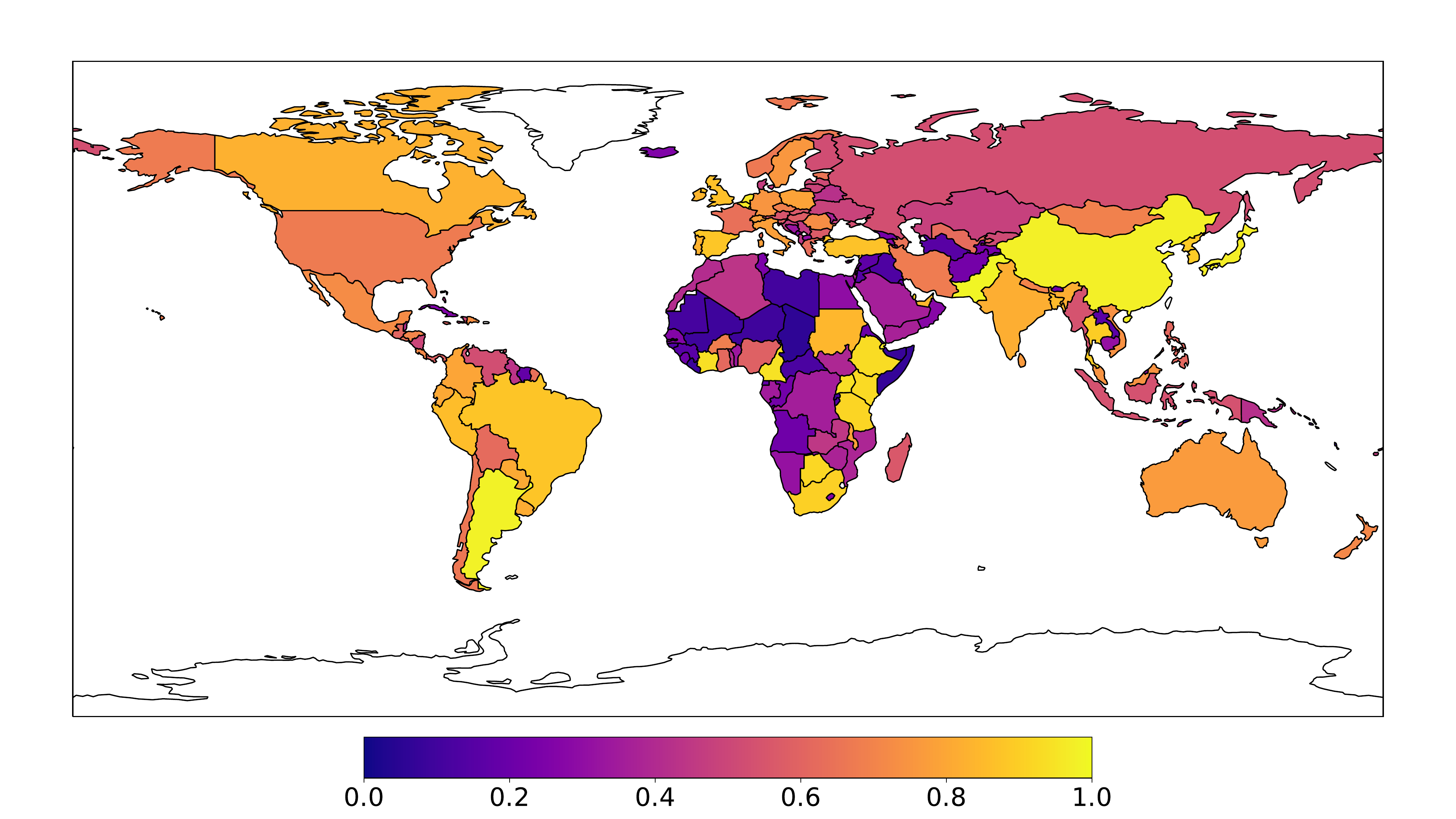}
\caption{Same as in Fig.~\ref{fig4}
for sensitivity to ICB China
with $D=4.75 \times 10^{-4}$ (blue color for zero)
and $D=5.91 \times 10^{-5}$ (yellow color for unity).
}
\label{fig6}
\end{center}
\end{figure}

The world influence of ICB China is shown in Fig.~\ref{fig6}.
The most sensitive country is Luxembourg ($D= 4.75 \times 10^{-4}$)
followed by Pakistan, Argentina, China, Japan, Djibouti.
Indeed, in 2011 ICB China opened a branch in 
Luxembourg which became its European headquarters
and its wiki-article has a direct link to  Luxembourg.
This article also directly points to
Pakistan where ICB China established its branches.
Argentina also has a direct link since ICB China acquired 
80\% of Standard Bank Argentina in 2012. 
The direct link to Japan results from the world's largest 
initial public offering (IPO)
which surpassed the previous record from Japan.
Djibouti has no direct links
but the wiki-article of Djibouti
points out that authorities have strengthened ties with China
that generates certain indirect links.
In global the world map of ICB China influence
is marked by its propagation to Africa and Europe.
The smallest sensitivity is for Chard ($D=5.91 \times 10^{-5}$).

The sensitivity world maps and 
the sensitivity values for all 195 countries and 
all 60 banks are available at \cite{wikibanknet}. 

\vspace{-0.4cm}
\section{Discussion}
\label{sec:7}

We performed a detailed analysis of interactions of the largest world banks 
\cite{wikibankrank,spreport} using the reduced Google matrix algorithm
and the network of English Wikipedia with more than 5 million articles.
While the top assets rank positions are occupied by the banks of China
with ICB China at the first position the Wikipedia network 
analysis shows that the most influential banks at the PageRank top positions
are US banks with the first one being Goldman Sachs bank.
The performed REGOMAX analysis allows to establish direct and 
hidden interactions 
between the largest world banks and countries
determining the closest friends and followers for each bank.
The sensitivity analysis gives the world map of countries
with their influence to a given bank.
This geographical analysis demonstrates a clear tendency
of  banks expansions to other countries like e.g. to East European 
countries for  
Deutsche Bank, African countries, Pakistan and Argentina for ICB China.
Since Wikipedia accumulates a huge amount of human knowledge
we argue that the Wikipedia network analysis captures the real
financial, social and historical interactions between 
world banks and countries.

We note that the financial networks of banks
have a relatively small size. Thus the whole Federal 
Reserve has only about 6000 nodes 
\cite{soramaki} that is about 1000 times smaller compared than the
Wikipedia network considered here. Thus we argue that the REGOMAX method
will allow to perform an efficient analysis of such financial networks
with new possibilities of crisis prevention. 

\vspace{-0.2cm}
\section*{Acknowledgments}

This work was supported in 
part by the Programme Investissements
d'Avenir ANR-11-IDEX-0002-02, 
reference ANR-10-LABX-0037-NEXT (project THETRACOM).
This work was granted access to the HPC GPU resources of 
CALMIP (Toulouse) under the allocation 2018-P0110. 
The development of the VexCL library was partially 
funded by the state assignment
to the Joint supercomputer center of 
the Russian Academy of Sciences for scientific
research.


\begin{thebibliography}{99}
\bibitem{fcrisiswiki} {\it Financial crisis of 2007 - 2008},
         \verb|https://en.wikipedia.org/w/index.php?title|\\
           \verb|=Financial\_crisis\_of\_2007|\\
          \verb|\%E2\%80\%932008&oldid=882711856|
          (Accessed February (2019)).
\bibitem{fcrisisguardian} {\it Three weeks that changed the world},
         The Guardian Dec 27 (2008),
         \verb|https://www.theguardian.com/business/2008|\\
          \verb|/dec/28/markets-credit-crunch-banking-2008|\\
           (Accessed February (2019)).
\bibitem{garratt2010} Bech M.L., Chapman J.T.E. and Garratt R.,
         {\it Which bank is the ``central'' bank?},
         J. Monetary Econ. {\bf 57} (2010) 352.
\bibitem{hale2012} Hale G.,
         {\it Bank relationships, business cycles, and financial crises},
         J. Int. Econ. {\bf 88} (2012) 312.
\bibitem{minoiu2013} Minoiu C. and Reyes J.A.,
         {\it A network analysis of global banking: 1978-2010},
         J. Fin. Stability {\bf 9} (2013) 168.
\bibitem{craig2014} Craig B. and von Peter G.,
         {\it Interbank tiering and money center banks},
          J. Finan. Intermediation {\bf 23(3)} (2014) 322.
\bibitem{wikibankrank} {\it List of largest banks}, Wikipedia,
           \verb|https://en.wikipedia.org/w/index.php?title=|
           \verb|List\_of\_largest\_banks\&oldid=871310902|
            (Accessed November (2018)).
\bibitem{spreport} {\it The world's 100 largest banks},
        S\&P Global Market Intelligence report, 6 April (2018),
        \verb|https://platform.mi.spglobal.com/web/client?|
        \verb|auth=inherit#news/article?id=44027195&cdid=|
        \verb|A-44027195-11060|  (Accessed February (2019)).
\bibitem{brin} Brin S. and Page L.,
         {\it The anatomy of a large-scale hypertextual Web search engine},
         Computer Networks and ISDN Systems {\bf 30} (1998) 107.
\bibitem{meyer} Langville A.M. and Meyer C.D., 
        {\it  Google's PageRank and beyond: 
         the science of search engine rankings}, 
        Princeton University Press, Princeton (2006).
\bibitem{rmp2015} Ermann L., Frahm K.M. and Shepelyansky D.L.,
        {\it Google matrix analysis of directed networks},
          Rev. Mod. Phys. {\bf 87} (2015) 1261.
\bibitem{wtn1} Ermann L. and Shepelyansky D.L.,
        {\it Google matrix of the world trade network},
        Acta Physica Polonica A {\bf 120} (2011) A158.
\bibitem{wtn2} Ermann L. and Shepelyansky D.L., 
         {\it Google matrix analysis of the multiproduct world trade network},
         Eur. Phys. J. B {\bf 88} (2015) 84.
\bibitem{gileswiki} Giles J.,
         {\it Internet encyclopaedias go head to head},
         Nature {\bf 438} (2005) 900.
\bibitem{reagle} Reagle J.M.Jr.,
         {\it Good faith collaboration: the culture of Wikipedia
          (History and Foundations of Information Science)},
         MIT Press (2012).
\bibitem{nielsen} Nielsen F.A.,
         {\it Wikipedia research and tools: review and comments},
         SSRN Electronic Journal (2012);
          Available \verb|https://doi.org/10.2139/ssrn.2129874|
           (Accessed February (2019)).
\bibitem{wikiquality} Lewoniewski W., Wecel K., Abramowicz W.,
          {\it  Relative quality and popularity
          evaluation of Multilingual Wikipedia Articles},
          Informatics {\bf 4(4)} (2017) 43.
\bibitem{24wiki2017} Frahm K.M. and Shepelyansky D.L.,
          {\it Wikipedia networks of 24 editions of 2017},
           \verb|http://www.quantware.ups-tlse.fr/QWLIB/|
          \verb|24wiki2017/| (2017) (Accessed February (2019)).
\bibitem{greduced} Frahm K.M. and Shepelyansky D.L.,
          {\it Reduced Google matrix},
          arXiv:1602.02394 [physics.soc] (2016).
\bibitem{politwiki} Frahm K.M., Jaffres-Runser K. and Shepelyansky D.L.,
         {\it Wikipedia mining of hidden links between political leaders},
          Eur. Phys. J. B {\bf 89} (2016) 269.
\bibitem{wikicountries} Frahm K.M., El Zant S., 
           Jaffres-Runser K. and Shepelyansky D.L.,
         {\it Multi-cultural Wikipedia mining of geopolitics interactions leveraging 
          reduced Google matrix analysis},
           Phys. Lett. A {\bf 381} (2017) 2677.
\bibitem{wrwu2017} Coquide C., Lages J. and Shepelyansky D.L., 
          {\it World influence and interactions of universities 
          from Wikipedia networks}, 
          Eur. Phys. J. B {\bf 92} (2019) 3.
\bibitem{proteins} Lages J., Shepelyansky D.L. and Zinovyev A., 
          {\it Inferring hidden causal relations between pathway members using 
            reduced Google matrix of directed biological networks},
          PLoS ONE {\bf 13(1)} (2018) e0190812.
\bibitem{wikibanknet} Demidov D., Frahm K.M. and Shepelyansky D.L.,
         {\it Wikipedia bank network},
         \verb|http://http://www.quantware.ups-tlse.fr/|
         \verb|QWLIB/wikibanknet/| (2019)
         (Accessed February (2019)).
\bibitem{cheirank}  Chepelianskii A.D.,
         {\it Towards physical laws for software architecture},
         arXiv:1003.5455 [cs.SE] (2010).
\bibitem{gitlabklaus} Frahm K.M. and Shepelyansky D.L.,
         {\it REGOMAX}, 
         \url{https://github.com/kmfrahm/regomax}  (Accessed February (2019)).
\bibitem{demidov2013} Demidov D., Ahnert K., Rupp K. and Gottschling P.,
    {\it Programming CUDA and OpenCL: A case study using modern C++ Libraries}.
    SIAM J. Sci. Comput. {\bf 35(5)} (2013) C453.
\bibitem{vexcl} Demidov D., VEXCL,
           \url{https://github.com/ddemidov/vexcl}  (Accessed December (2018)).
\bibitem{cuthillmckee} Cuthill E. and McKee J., {\it Reducing the
   bandwidth of sparse symmetric matrices.} In Proc. 24th Nat. Conf.
   ACM,  p.157 (1969).
\bibitem{olympe} Olympe, CALMIP 
        \url{https://www.calmip.univ-toulouse.fr/spip.php?article582} 
               (Accessed December (2018)).
\bibitem{gitlabdenis} Demidov D., {\it REGOMAX-GPU},
    \url{https://gitlab.com/ddemidov/regomax-gpu} (Accessed February (2019)).
\bibitem{ieeedis} Rollin G., Lages J. and Shepelyansky D.L.,
         {\it World influence of infectious diseases from Wikipedia network analysis},
          IEEE Access doi.org/10.1101/424465 (2018).
\bibitem{dorogovtsev} Dorogovtsev S., \textit{Lectures on complex networks},
          Oxford University Press, Oxford (2010)
\bibitem{goldmanwiki} {\it Goldman Sachs},
         \verb|https://en.wikipedia.org/wiki/|
          \verb|Goldman_Sachs#cite_note-34|
           (Accessed February (2019)).
\bibitem{soramaki} Soramaki K., Bech M.L., Arnold J., Glass R.J. and Beyeler W.E.,
        {\it The topology of interbank payment flows},
         Physica A {\bf 379} (2007) 317.

\end{thebibliography}
\end{document}